\documentclass[12pt]{article}
\usepackage{graphicx}
\usepackage{latexsym, pstricks}
\usepackage{amsmath,amssymb,amsthm}
\usepackage[labelformat=empty,labelsep=none]{caption}

\font\titulo=cmbx10 scaled\magstep1 
scaled\magstep1

\def\section#1{\vskip 1.5truepc plus 0.1truepc minus 0.1truepc
    \goodbreak \leftline{\titulo#1} \nobreak \vskip 0.1truepc
    \indent}
\def\frc#1#2{\leavevmode\kern.1em
    \raise.5ex\hbox{\the\scriptfont0 $ #1 $}\kern-.1em
    /\kern-.15em\lower.25ex\hbox{\the\scriptfont0 $ #2 $}}

\def\IZ{{\rm Z}\llap{\vrule height7.1pt width1pt
     depth-.4pt\phantom t}} 

\newbox\pmbbox
 \def\pmb#1{{\setbox\pmbbox=\hbox{$#1$}%
\copy\pmbbox\kern-\wd\pmbbox\kern.3pt\raise.3pt\copy\pmbbox\kern-\wd\pmbbox
\kern.3pt\box\pmbbox}}

\font\cmss=cmss10 \font\cmsss=cmss10 at 7pt

\def\IZ{\relax\ifmmode\mathchoice
{\hbox{\cmss Z\kern-.4em Z}}{\hbox{\cmss Z\kern-.4em Z}}
{\lower.9pt\hbox{\cmsss Z\kern-.4em Z}} {\lower1.2pt\hbox{\cmsss
Z\kern-.4em Z}}\else{\cmss Z\kern-.4em Z}\fi}

\font\cmss=cmss10 \font\cmsss=cmss10 at 7pt
\def\IS{\relax\ifmmode\mathchoice
{\hbox{\cmss S\kern-.4em S}}{\hbox{\cmss S\kern-.4em S}}
{\lower.9pt\hbox{\cmsss S\kern-.4em S}} {\lower1.2pt\hbox{\cmsss
S\kern-.4em S}}\else{\cmss S\kern-.4em S}\fi}

\begin{document}

\centerline{\titulo A Parametric Bootstrap for the Mean Measure of Divergence}

\

\centerline{Zertuche, F.${}^{a*}$ and Meza-Pe\~naloza, A.${}^b$}

\

\leftline{${}^{a}$ U. Cuernavaca, Instituto de Matem\'aticas, Universidad Nacional Aut\'onoma}
\leftline{de M\'exico, Mexico}

\

\leftline{${}^b$ Instituto de Investigaciones Antropol\'ogicas, Universidad Nacional Aut\'onoma}
\leftline{de M\'exico, Mexico}

\vskip 1.0pc {\bf \centerline {Abstract}}

For more than $50$ years the {\it Mean Measure of Divergence} (MMD) has been one of the most prominent tools used in anthropology for the study of non-metric traits. However, one of the problems, in anthropology including palaeoanthropology (more often there), is the lack of big enough samples or the existence of samples without sufficiently measured traits. Since 1969, with the advent of bootstrapping techniques, this issue has been tackled successfully in many different ways. Here, we present a parametric bootstrap technique based on the fact that the transformed $ \theta $, obtained from the Anscombe transformation to stabilize the variance, nearly follows a normal distribution with zero mean and variance $ \sigma^2 = 1 / (N + 1/2) $, where $ N $ is the size of the measured trait. When the probabilistic distribution is known, parametric procedures offer more powerful results than non-parametric ones. We profit from knowing the probabilistic distribution of $ \theta $ to develop a parametric bootstrapping method. We explain it carefully with mathematical support. We give examples, both with artificial data and with real ones. Our results show that this parametric bootstrap procedure is a powerful tool to study samples with scarcity of data.

\vskip 0.5pc

\noindent {\bf Short title:} {\it A Parametric Bootstrap}



\vskip 0.5pc

\leftline{${}^*$Corresponding author.}
\leftline{E-mail address: \texttt{federico.zertuche@im.unam.mx} (F. Zertuche)}

\newpage

\baselineskip = 24.8pt

\section{1. Introduction}

In the anthropological literature, cranial non-metric traits are which is called categorical binary variables in the mathematical literature. They are usually recorded as present or absent. Around $200$ variables have been described for the skull (Le Double, 1903; Ossenberg, 1976; Hauser, \& De Stefano, 1989; Pink, Maier, Pilloud, \& Hefner, 2016), and a similar number have been identified for the post-cranial skeleton (Finnegan, 1978; Donlon, 2000; Voisin, 2012; Verna, {\it et al.} 2013). These traits should not be confused with the {\it macromorphoscopic} traits, such as bone shape, bone feature shape, or sutural shape (Hefner, \& Linde, 2018). Several researches have studied the reliability of using non-metric traits in studies of biological affinity. Their results have shown that there is a strong genetic component linked to the presence and transmission of these traits at the phenotypic level (Sj{\o}vold, 1977; McGrath, Cheverud, \&  Buikstra, 1984; Pink, {\it et al.}, 2016).

For more than $50$ years, since its introduction by C.A.B. Smith for use by Grewal (1962) to establish genetics distances between different strains of mice, the {\it Mean Measure of Divergence} (MMD) has been widely used after its introduction in the field of anthropology as a way to study the separation between populations through non-metric traits by Berry, \& Berry, (1967). After some improvements over the years, the MMD turns out to be of wide use in anthropology when non-metric traits are used (Green, \& Suchey, 1976; Rothhammer, Quevedo, Cocilovo, \& Llop, 1984; Hanihara, Ishida, \& Dodo, 2003; Sutter, \& Mertz, 2004; Schillaci, Irish, \& Wood, 2009; Williams, \& Cofran, 2016). Furthermore, the use of non-metric traits and the MMD has been spread to other scientific field such as biology to study fauna (Sikorski,1982; Ansorge, 2001; Ansorge, Ranyuk, Kauhala, Kowalczyk \& Stier, 2009), and even in informatics (Suryakan, \& Mahara, 2016). All of these issues turn the MMD a preponderant measure of separation between populations (Hartman,1980; Kry\v{s}tufek,1990; Trimble \& Praderi 2008; Markov {\it et al.} 2018).

Among other things, the MMD biodistance has good statistical properties: it is a statistical unbiased estimator of its real value, and its statistical significance level $ \alpha $ can be easily evaluated (de Souza, \& Houghton, 1977).

Let $ M $ denote the number of non-metric traits under consideration, which we enumerate by the Latin subindex $ i $ ($ 1 \leq i \leq M $). Also, let $ P $ denote the number of populations that we are dealing with. We denote them by the Greek subindexes $ \mu $, $ \nu = 1, 2, \ldots, P $. Let $ N_\mu $ be the size of the $\mu$-population. Non-metric traits are categorical variables that we denote by `0' (absent) and `1' (present). When the trait was unable be measured we add a `?' in the data base and no count is done. For each $ \mu $-population and each $ i $-trait, let $ K_{\mu i} $ be the number of times the $ i $-trait is present. We denote the number of measures of the $ i $-trait by $ N_{\mu i} $. Due to these definitions, $ K_{\mu i} \leq N_{\mu i} \leq N_\mu $, because it could happen that it was not possible to measure all of the traits.

Under the {\it null hypothesis of no causation}, each non-metric trait should follow a binomial distribution with $ N_{\mu i} $ events, mean $ p_{\mu i} = K_{\mu i}  / N_{\mu i} $, and variance $ \varrho_{\mu i}^2 = p_{\mu i} \ (1 - p_{\mu i}) / N_{\mu i} $ (Wackerly, Mendenhall, \& Scheaffer, 2008). The idea first introduced by Bartlett (1947), later improved by Anscombe (1948), and Freeman, Tukey (1950) was to stabilize the variance through an inverse trigonometric transformation. Both Anscombe and Freeman-Tukey transformations give almost the same results even for small samples with the best performance (Nikita 2015). In what follows we are going to use Anscombe's transformation which is given by:
$$
\theta_{\mu i} \ = \ \sin^{-1} \left(1-2 \ {K_{\mu i} + 3/8 \over N_{\mu i} + 3/4} \right) \ ,
$$
where $ \theta_{\mu i} $ is measured in radians ($ - \pi / 2 < \theta_{\mu i} < \pi / 2 $). Under the {\it null hypothesis} $ \theta_{\mu i} $ follows nearly a normal distribution with zero mean and standard deviation
$$
\sigma_{\mu i} \ = \ {1 \over \sqrt{N_{\mu i} + 1/2}}     \eqno(1)
$$
(de Souza, \& Houghton, 1977). For short, we are going to denote this by $ {\cal N} \left(0, \, \sigma_{\mu i} \right) $.

The MMD biodistance between two populations $ \mu $ and $ \nu $ is given by (de Souza, \& Houghton, 1977; Nikita, 2015)
$$
\textrm{MMD}\left(\mu,\ \nu \right) \ = \ {1 \over M}\ \sum_{i=1}^M \ \left\{ \left( \theta_{\mu i} - \theta_{\nu i} \right)^2 - {1 \over N_{\mu i} + 1/2}  - {1 \over N_{\nu i} + 1/2} \right\} \ .
$$
Its variance under the {\it null hypothesis} was also calculated by de Souza, \& Houghton (1977), who  obtained
$$
\sigma^2 \left(\mu,\ \nu \right) \ = \ {2 \over M^2} \sum_{i=1}^{M}{{\ \ \left\{\frac{1}{N_{\mu i}\ + 1/2} + \ \frac{1}{N_{\nu i}\ + 1/2 }\right\}}^2\ } \ .
$$
If $ \textrm{MMD}\left(\mu,\ \nu \right) \ \leq \ 0 $, the $ \textrm{MMD}\left(\mu,\ \nu \right) $ should be set to zero, meaning that the two populations are indistinguishable. A big note of {\bf caution}: This could happen because they are really close populations or because there are few samples in one or both populations. Is is important to note that even if $ \textrm{MMD}\left(\mu,\ \nu \right) \ > \ 0 $, this may not imply a representative biodistance. A representative MMD biodistance should be bigger than the statistical fluctuations. To address this situation, it is better to work with the standardized MMS biodistance defined by:
$$
\textrm{stMMD}(\mu, \nu) \ = \ \textrm{MMD}(\mu, \nu) \ / \ \sigma \left(\mu,\ \nu \right) \ . \eqno(2)
$$
The stMMD biodistance has the great advantage that, under the {\it null hypothesis}, it nearly follows an $ {\cal N} \left(0, \, 1 \right) $ distribution (de Souza, \& Houghton, 1977). This may be used to test statistical representativeness. In this case, if $ \textrm{stMMD}(\mu, \nu) \ > \ 2 $, its significance level will be better than $ \alpha = 0.05 $. In the rest of this work, all biodistances will be expressed by (2). If the reader wants a more accurate statistical test of representability, the quantity
$$
\chi^2 \ = \ \sum_{i=1}^M \ {\left( \theta_{\mu i} - \theta_{\nu i} \right)^2 \over \left( N_{\mu i} + 1/2 \right)^{-1} + \left( N_{\nu i} + 1/2 \right)^{-1}}   \ ,      \eqno(3)
$$
follows by construction a $ \chi^2_M $ probabilistic distribution under the {\it null hypothesis}.

\newpage

\section{2. A parametric bootstrap procedure}

With the advent of fast computers, bootstrapping procedures have grown (Efron 1979). The basic idea is to re-sample populations with replacement, allowing for a better representability of the small ones. These methods are widely used by anthropologists and paleoanthropologists when samples are scarce (Irish, Guatelli-Steinberg, Legge, de Ruiter, Berger, 2013; Movsesian, 2013; Carter, Worthington, \& Smith, 2014; Movsesian, Bakholdina, \& Pezhemsky, 2014; Villmoare, 2005; Schillaci, {\it et al.}, 2009).

When the collected data follow a known probabilistic distribution, parametric procedures are more accurate. Nevertheless, one must be sure about the probabilistic distribution involved, or have a good guess of it, in order for them to work properly. Fortunately, this is our case, since $ \theta_{\mu i} $ follows an $ {\cal N} \left(0, \, \sigma_{\mu i} \right) $ with $ \sigma_{\mu i} $ given by (1). Our parametric bootstrap procedure consists in re-sampling the standard deviation $ \sigma_{\mu i} $ through $ {\cal N} \left(0, \, \sigma_{\mu i} \right) $. It is important to note that, if the bootstrap procedure is going to be applied, it should be done in all populations and each non-metric trait, regardless their degree of representativeness.

Let us be specific about this parametric bootstrap procedure: For each $ \mu $-population  and $ i $-trait, $ N_{\mu i} $ samples $ \hat{\theta}_{\mu i}^{(s)} $ ($ 1 \leq s \leq N_{\mu i} $) are collected from a normal distribution with zero mean and standard deviation (1), {\it i.e.} $ {\cal N} \left(0, \sigma_{\mu i} \right) $. Through the set $ \left\{ \hat{\theta}_{\mu i}^{(s)} \right\} $, we calculate a new standard deviation $ \sigma_{\mu i}^* $ in the standard way {\it i.e.}
$$
\sigma_{\mu i}^* \ = \ \sqrt{{1 \over N_{\mu i} - 1} \ \sum_{s=1}^{N_{\mu i}} \ \left\{ \hat{\theta}_{\mu i}^{(s)} - \bar{\theta}_{\mu i} \right\}^2}     \eqno(4)
$$
where
$$
\bar{\theta}_{\mu i} \ = \ {1 \over N_{\mu i}} \ \sum_{s=1}^{N_{\mu i}} \ \hat{\theta}_{\mu i}^{(s)} \eqno(5)
$$
is the average.

We repeat this procedure $ B $ times by doing new extractions \break from $ {\cal N} \left(0, \sigma_{\mu i} \right) $ and calculating $ \sigma_{\mu i}^* $, once again through (4) and (5). Let us add a $ \left( b \right) $-superindex to each of these procedures. At the end, we obtain the set
$$
\Xi_{\mu i}^{*(B)} = \left\{ \sigma_{\mu i}^{* (1)}, \sigma_{\mu i}^{* (2)}, \ldots, \sigma_{\mu i}^{* (b)}, \ldots, \sigma_{\mu i}^{* (B)} \right\} \ ,  \eqno(6)
$$
using (4) and (5) in all the calculations. The bootstrapped value $ \mathfrak{S}_{\mu i} $ of $ \sigma_{\mu i} $ is then given by
$$
\mathfrak{S}_{\mu i} \ = \ \sqrt{{1 \over B - 1} \ \sum_{b=1}^B \ \left\{ \sigma_{\mu i}^{* (b)} - \bar{\sigma}_{\mu i} \right\}^2}   \ ,  \eqno(7)
$$
where
$$
\bar{\sigma}_{\mu i} \ = \ {1 \over B} \ \sum_{b=1}^B \ \sigma_{\mu i}^{* (b)} \ .
$$
Since $ B \gg 1 $, the denominator $ B - 1 $ inside the square root of (7) may well be substituted by $ B $ without a significant change in the results. In general, values of $ B = 500 $ are enough, and results do not change significantly for a bigger $ B $.

Now, a bootstrapped $ \textrm{MMD}_B $ biodistance may be calculated through
$$
\textrm{MMD}_B \left(\mu,\ \nu \right) \ = \ {1 \over M}\ \sum_{i=1}^M \ \left\{ \left( \theta_{\mu i} - \theta_{\nu i} \right)^2 - \mathfrak{S}_{\mu i}^2 - \mathfrak{S}_{\nu i}^2 \right\} \ ,
$$
with its corresponding standard deviation
$$
\sigma_B \left(\mu,\ \nu \right) \ = \ \sqrt{{2 \over M^2} \sum_{i=1}^{M}{{\ \left\{ \mathfrak{S}_{\mu i}^2 \ + \ \mathfrak{S}_{\nu i}^2 \right\}}^2\ } }  \ ,
$$
and the standardized $ \textrm{MMD}_B $
$$
\textrm{stMMD}_B \left(\mu, \nu \right) = \textrm{MMD}_B \left(\mu, \nu \right) \ / \ \sigma_B \left(\mu,\ \nu \right) \ .
$$

\

\section{3. Practical considerations}

All programs were implemented by programming in \textmd{C}, the use of \break {\it GNUPLOT}~5.0 graphics program and {\it PAST}~3.21. The algorithm may be implemented directly in {\it PAST}, {\it R}, or other statistical packages using their script programming by easy extensions of our approach.

It is important to have a good uniform pseudo-random number generator on the open interval $ \left( 0, 1 \right) $: which is the base for constructing a Gaussian generator by means of the {\it Box-Muller} method (Box, \& Muller, 1958). The one we used has a period $ > 2 \times 10^{18} $ (Press, Teukolsky, Vetterling, \& Flannery, 1996). With these measures, the reader would be able to have Gaussian pseudo-random deviates so good to pass the {\it Shapiro-Wilk} test of normality with $ p \sim 0.9\ldots $ (Shapiro, \& Wilk, 1965).

\

\section{4. How it works}

For simplicity, let us focus in just one variable $ N $, being the measured size of a population's non-metric trait. We will eliminate all the subindexes in the rest of this section. Because of (1) we have
$$
\sigma \ = \ {1 \over \sqrt{N + 1/2}}   \ ,  \eqno(8)
$$
In Fig.~1, the behavior of $ \sigma $ and the bootstrapped $ \mathfrak{S} $ obtained from (7) are shown {\it vs.} $ N $. The reader can see that as the sample's size $ N $ increases, the value of $ \sigma $ decreases following (8), and the parametric bootstrap always gives a smaller $ \mathfrak{S} $. We did a least squares fit of $ \mathfrak{S} $  {\it vs.} $ N $ by previously doing a log-log transformation. The resulting $ \mathfrak{S}_{ls} $ is shown in Fig.~2, and it is given by
$$
\mathfrak{S}_{ls} \ = \ {A \ \over \ N^\beta} \ ,   \eqno(9)
$$
where $ A = 0.389797 $, $ \beta = 1.0028 $, and a correlation $ r = -0.99532 $ with a confidence level of $ \alpha \sim 10^{-99} $. It is not surprising that $ \beta \sim 1 $ due to the {\it Central Limit Theorem} (Bulmer, 1979; Wackerly, {\it et al.} 2008). To show this, let \emph{big} $ {\cal O} \left( f(x) \right) $ represents the asymptotic behavior of a function $ f(x) $ where only the biggest term without constant factors is taken into account as $ x \gg 1 $. In the asymptotic regime $ N \gg 1 $,  $ \sigma^* $  behaves like
$$
\sigma^* \ \approx \ {\cal O} \left( 1 /  \sqrt{N} \right) \, \sigma \ .
$$
From (8) $ \sigma \approx \mathcal{O} \left( 1 / \sqrt{N} \right) $, so it follows that $ \sigma^* \approx \mathcal{O} \left( 1 / N \right) $. This is the order of magnitude for all the elements in (6) {\it i.e.} $ \sigma^{*(b)} \in \Xi^{*(B)} $. Since all of them are uncorrelated, from (7) follows that
$$
\mathfrak{S} \ \approx \ {\cal O} \left( {1 \over N} \right) \ ,
$$
explaining why $ \beta \sim 1 $ in (9).

This is the general view for one variable. Nevertheless, when all the $ M $ non-metric traits are taken into consideration, this behavior becomes milder and more complex, since each $i$-trait for each $\mu$-population has a different $ N_{\mu i} $. This becomes specially apparent when there is a lack of statistical representability.

\

\section{5. How it should be applied and interpreted}

On the basis of the foregoing section, it is clear that small populations are the ones with more correction in $ \sigma_{\mu i} $, and so their stMMD biodistance with respect to other populations will grow more. This is a nice thing, since small populations have less statistical representation. However, it can happen that a $ \mu $-population with a big $ N_\mu $ does not have a big $ N_{\mu i} $ for every $i$. In that instance, it will not have a good statistical representability, and the parametric bootstrap will greatly correct $ \sigma_{\mu i} $, as is the case with small populations. The quantity
$$
\mathfrak{T}_\mu \ = \ {1 \over N_\mu} \ \sum_{i = 1}^M \ N_{\mu i} \ ,  \eqno(10)
$$
measures the percentage of information of the $\mu$-population. When $ \mathfrak{T}_\mu < 0.5 $, one might expect the $\mu$-population not to have a good statistical representability.

The big picture is as follows:

When all the populations have good statistical representation, the bootstrapping procedure will change the stMMD distances matrix by {\it almost} a re-scaling factor $ \lambda $. By {\it almost}, we understand that, while the factor may be different for each entry of the distances matrix, the ranking order of the distances is preserved. In this way, if we applied ordering algorithms like {\it non-metric multidimensional scaling} (nMDS) or clustering algorithms like {\it Unweighted Pair Group Method with Arithmetic mean} (UPGMA), results will not change. A way to see this is the following: Let $ \mathfrak{D}^* $ and $ \mathfrak{D} $ be the distances matrices of the $ P $ populations with and without bootstrap, respectively. In the ideal case of a re-scaling, we should have
$$
\mathfrak{D}^* \ = \ \lambda \ \ \mathfrak{D}    \ .       \eqno(11)
$$
In general, this is not the case, but the quotients
$$
\mathfrak{C}_{\mu \nu} = \mathfrak{D}_{\mu \nu} / \mathfrak{D}^*_{\mu \nu}
$$
of the matrices entries give a statistics of what is happening. Note that we define $ \mathfrak{C}_{\mu \nu} $ by placing $ \mathfrak{D}^*_{\mu \nu} $ in the denominator. We do this since if the parametric bootstrap procedure is working well, $ \mathfrak{D}^*_{\mu \nu} \neq 0 $ for all $ \mu $, $ \nu $ such that $ \mu \neq \nu $. On the contrary, some of the $ \mathfrak{D}_{\mu \nu} $ entries, with $ \mu \neq \nu $, could well be zero, specially in the case of statistically misrepresented populations.

Since the number of distances $ N_d $ between $ P $ populations is \break $ N_d = P (P - 1) / 2 $, the mean and the standard deviation of $ \lambda^{-1} $ are given by
$$
\left< \lambda^{-1} \right> \ = \ {1 \over N_d} \ \sum_{\mu = 1}^P \ \sum_{\nu = \mu + 1}^P \, \mathfrak{C}_{\mu \nu}
$$
and
$$
\mathfrak{R} \ = \ \sqrt{{1 \over N_d - 1} \ \sum_{\mu = 1}^P \ \sum_{\nu = \mu + 1}^P \, \left\{ \mathfrak{C}_{\mu \nu} - \left< \lambda^{-1} \right> \right\}^2} \ ,
$$
respectively. The relative error (in percentage) of $ \lambda^{-1} $ is given by
$$
{\cal E} \ = \ {100 \ \times \ \mathfrak{R} \ \over \ \left< \lambda^{-1} \right>}  \ . \eqno(12)
$$
When $ {\cal E} \lesssim 10 \% $, there is a confidence that the populations have statistical representability. Instead, if $ {\cal E} > 10 \% $, the rank order in the distances matrix may have already changed. This will happen when one, or some, of the populations lack statistical representability. Therefore, since the rank order in the distances matrix has changed, this would be reflected in the nMDS and UPGMA algorithms.

To illustrate these points, in the next sections we will apply the parametric bootstrap algorithm first to populations obtained from artificial data, and then to real populations.


\

\section{6. Artificial data}

We generate artificial data for $ 10 $ populations with $ M = 13 $ non-metric traits. A random generator of uniform numbers creates each population with a maximum size of $ N = 100 $. We include two thresholds $ \Lambda_1 $ and $ \Lambda_2 $. $ \Lambda_1 $ establishes the minimum size of the populations and $ \Lambda_2 $ the maximum tolerance of non-measured traits. In this way, we have control of the populations' statistical representativeness.

\

The first artificial data correspond to populations with a good statistical representativeness. In this case, we set $ \Lambda_1 = 50 $ and $ \Lambda_2 = 45 $. The details of the populations are shown in Table~S1 (supplementary material). Also the distances matrices, without bootstrap and with bootstrap, are shown in Tables~S2 and S3 respectively. From them we calculate the relative error (12), obtaining the value $ {\cal E} \ = \ 5.9881 \% $. This result indicates that the populations have a good statistical representativeness. This fact can also be corroborated in Table~1. There, we see that the net size of all populations is big, and also that all $ \mathfrak{T}_\mu > 0.5 $. In Fig.~3, a dendrogram with UPGMA is shown for the distances matrix S2, where no bootstrap was done. In Fig.~4, the corresponding dendrogram for the distances matrix S3, where the parametric bootstrap algorithm was applied, is shown. We see that, up to {\it almost} a re-scaling factor, the two dendrograms are the same as expected.

\footnotesize
\begin{center}
\begin{tabular}{|c|c|c|c|c|c|c|c|c|c|c|}
\hline $ \textrm{Population} \ \mu \rightarrow $ & {\bf 1} & {\bf 2} & {\bf 3} & {\bf 4} & {\bf 5} & {\bf 6} & {\bf 7} & {\bf 8} & {\bf 9} & {\bf 10} \\
\hline $ N_\mu \longrightarrow  $ & 93 & 76 & 68 & 58 & 64 & 64 & 93 & 62 & 72 & 85  \\
\hline $ \mathfrak{T}_\mu \longrightarrow  $ & 0.78& 0.82& 0.83& 0.91& 0.88& 0.86& 0.74& 0.86& 0.87& 0.69\\
\hline

\end{tabular}
\end{center}
\noindent
\textbf{Table~1.} The ten populations of the first artificial data with their sizes and percentages of information as given by (10).

\baselineskip = 24.8pt

\normalsize

\baselineskip = 24.8pt

\

The second artificial data correspond to populations with a bad statistical representativeness. For that scope, we set $ \Lambda_1 = \Lambda_2 = 2 $. The details of the populations are shown in Table~S4 (supplementary material). The distances matrices, without bootstrap and with bootstrap, are shown in Tables~S5 and S6, respectively. In contraposition to the first artificial data, the relative error (12) has the value $ {\cal E} \ = \ 39.26 \% $, which is an indication that the populations have bad statistical representativeness. In Table~2, we see that the net size of all populations varies from just $ 4 $ samples to $ 99 $. Also, some populations have $ \mathfrak{T}_\mu < 0.5 $ values, even though their net size is big: see for example the case of the first population.

\footnotesize
\begin{center}
\begin{tabular}{|c|c|c|c|c|c|c|c|c|c|c|}
\hline $ \textrm{Population} \ \mu \rightarrow $ & {\bf 1} & {\bf 2} & {\bf 3} & {\bf 4} & {\bf 5} & {\bf 6} & {\bf 7} & {\bf 8} & {\bf 9} & {\bf 10} \\
\hline $ N_\mu \longrightarrow  $ & 46& 76& 4& 59& 99& 5& 66& 36& 88& 38\\
\hline $ \mathfrak{T}_\mu \longrightarrow  $ & 0.35& 0.62& 0.69& 0.48& 0.51& 0.71& 0.39& 0.54& 0.58& 0.61\\
\hline

\end{tabular}
\end{center}
\noindent
\textbf{Table~2.} The ten populations of the second artificial data with their sizes and percentages of information as given by (10).

\baselineskip = 24.8pt

\normalsize

\baselineskip = 24.8pt

In Fig.~5, a dendrogram with UPGMA is shown for the distances matrix S5, where no bootstrap was done. In Fig.~6, the corresponding dendrogram for the distances matrix S6, where the parametric bootstrap was applied, is shown. We can now observe that the two dendrograms are completely different, as was expected. In this case, the parametric bootstrap procedure should be taken as the accurate case for solving the lack of statistical representativeness.


\section{7. Real data}

In Meza-Pe\~naloza, \& Zertuche, (2019), we presented a work in which this parametric bootstrap procedure was developed and used. The seven populations studied were: Tlatilco ($ \mu =1 $), Teotihuacan ($ \mu = 2 $), {\it Epiclassic} Xico ($ \mu =3 $), Toluca valley ($ \mu =4 $), Xaltocan ($ \mu =5 $), Mogotes ($ \mu =6 $), and {\it Postclassic} Xico ($ \mu =7 $). The have been ordered almost chronologically. In Table~S7 (supplementary material), the data base of the populations may by consulted. In Table~3 the statistical data of the populations are shown.

\footnotesize
\begin{center}
\begin{tabular}{|c|c|c|c|c|c|c|c|}
\hline $ \textrm{Population} \ \mu \rightarrow $ & {\bf 1} & {\bf 2} & {\bf 3} & {\bf 4} & {\bf 5} & {\bf 6} & {\bf 7} \\
\hline $ N_\mu \longrightarrow  $ & 78& 66& 5& 23& 118& 15& 28\\
\hline $ \mathfrak{T}_\mu \longrightarrow  $ & 0.93& 0.31& 0.92& 0.95& 0.82& 0.71& 0.88\\
\hline

\end{tabular}
\end{center}
\noindent
\textbf{Table~3.} The seven populations extracted from real data with their sizes and percentages of information as given by (10). Tlatilco ($ \mu =1 $), Teotihuacan ($ \mu = 2 $), {\it Epiclassic} Xico ($ \mu =3 $), Toluca valley ($ \mu =4 $), Xaltocan ($ \mu =5 $), Mogotes ($ \mu =6 $), and {\it Postclassic} Xico ($ \mu =7 $).

\normalsize

\baselineskip = 24.8pt

As can be seen, one of the populations ({\it Epiclassic} Xico) have just $ 5 $ individuals. On the other hand, Teotihuacan, with $ 66 $ individuals, has a low percentage of information ($ \mathfrak{T}_3 = 0.31 $). Therefore, we would expect that the parametric bootstrap procedure can be applied. The distances matrix for the stMMD without bootstrap may be found in Table~S8, and the distances matrix with parametric bootstrap in Table~S9. From them we calculate the relative error (12), obtaining the value $ {\cal E} = 95.29 \% $. So, no doubt that things will change with the parametric bootstrap. The dendrogram obtained without bootstrap is shown in Fig.~7, whereas the one with the parametric bootstrap appears in Fig.~8. In the latter case, note that the smallest distance between {\it Epiclassic} Xico and {\it Postclassic} Xico is statistically representative since stMMD = 14.23, and from (3) $ \chi^2_{13} = 101.1234 $, giving a level of significance $ \alpha < 10^{-17} $.

\

\newpage

\section{8. Conclusion}

We have developed a parametric bootstrap procedure for the MMD biodistance. It is described in detail in Sec.~2. It can be applied with confidence in any set of populations. When they have a good statistical representativeness, all the distances will be almost re-scaled by a $ \lambda $ factor defined in (11), and a relative error (12) of the order $ {\cal E} \lesssim 10 \% $. Thus, the ranking order of the distances is going to be preserved, and nor will the results with the UPGMA cluster analysis change, nor those with the nMDS ordination method. On the contrary, when $ {\cal E} > 10 \% $, we have an indication of a bad statistical representativeness, therefore the distances ranking order begins to change. In this case, the parametric bootstrap is advisable. A lack of statistical representativeness may be detected when some of the populations have less than 10 individuals. It may also happen that $ N_\mu $ is big, but the percentage of information (10) is $ \mathfrak{T}_\mu < 0.5 $.

\

\centerline{\bf Acknowledgments}

The authors thank \textsf{Pilar L\'opez Rico} for her informatics services; \textsf{Rub\'en Gonz\'alez Zainez}, and \textsf{Patricia Pel\'aez Flores} for their computer advise.

\

\

\

\centerline{{\bf Literature Cited}}

\begin{itemize}

\item[] Anscombe F. J. (1948). The transformation of Poisson, binomial, and negative-binomial data. Biometrika, 35(3-4), 246-254.

\item[] Ansorge H. (2001). Assessing non-metric skeleton characters as a morphological tool. Zoology, 104, 268-277.

\item[] Ansorge H., Ranyuk M., Kauhala K., Kowalczyk R., \& Stier N. (2009). Raccoon dog, Nyctereutes procyonoides, populations in the area of origin and in colonised regions - the epigenetic variability of an immigrant. Ann. Zool. Fennici 46, 51-62.

\item[] Bartlett M. S. (1947). The Use of Transformations. Biometrics, 3(1), 39-52.

\item[] Berry, A. C., \& Berry R. J. (1967). Epigenetic variation in the human cranium. J~Anat, 101(2), 361-379.

\item[] Box, G. E. P., \& Muller, M. E. (1958). A Note on the Generation of Random Normal Deviates. Ann~Math~Statist, 29(2), 610-611.

\item[] Bulmer M. G. (1979). Principles of Statistics. Dover New York.

\newpage

\item[] Carter, K., Worthington, S., \& Smith, T. M. (2014). News and views: Non-metric dental traits and hominin phylogeny. J~Hum~Evol, 69, 123-128.

\item[] Cheverud, J. M., \& Buikstra J. E. (1981). Quantitative genetics of skeletal nonmetric traits in the rhesus macaques on Cayo Santiago. II. Phenotypic, genetic, and environmental correlations between traits. Am~J~Phys~Anthropol, 54(1), 51-58.

\item[] Donlon, D.A. (2000). The value of infracranial nonmetric variation in studies of modern Homo sapiens: an Australian focus. Am~J~Phys~Anthropol, 113(3), 349-368.

\item[] Efron, B. (1979). Bootstrap methods: Another look at the jackknife. Ann~Statist, 7(1), 1-26.

\item[] Finnegan, M. (1978). Non-metric variation of the infracranial skeleton. J~Anat, 125(Pt1), 23.

\item[] Freeman, M.F., Tukey, J.W. (1950). Transformations related to the angular and square root. Ann~Math~Stat, 21, 607-611.

\item[] Green, R.F., \& Suchey, J. M. (1976). The use of inverse sine transformations in the analysis of non-metric cranial data. Am~J~Phys~Anthropol, 45, 61-68.

\item[] Grewal, M. S. (1962). The rate of genetic divergence in the C57BL strain of mice. Genet~Res, 3(2), 226-237.

\item[] Hanihara, T., Ishida, H., \& Dodo, Y. (2003). Characterization of biological diversity through analysis of discrete cranial traits. Am~J~Phys~Anthropol, 121(3), 241-251.
\item[]Hartman, S. E. (1980). Geographic variation analysis of Dipodomys ordii using nonmetric cranial traits. Journal of Mammalogy, 61(3), 436-448.

\item[] Hauser, G., \& De Stefano, G. D. (1989). Epigenetic variants of the human skull (Ed.). Schweizerbart Science Publishers.

\item[] Hefner, J. T., \& Linde, K. C. (2018). Atlas of Human Cranial Macromorphoscopic Traits. Academic Press.

\item[] Irish, J. D., Guatelli-Steinberg, D., Scott, S., Legge, S. S., de Ruiter, D. J., \& Berger, L. R. (2013). Dental Morphology and the Phylogenetic ``Place" of Australopithecus sediba. Science, 340, 1233062-4.

\item[] Le Double, A.F. (1903). Trait\'e des variations des os du cr\^{a}ne de l'homme et de leur signification au point de vue de l'anthropologie zoologique. Vigot.

\item[] Kry\v{s}tufek, B. (1990). Nonmetric cranial variation and divergence of European sousliks (Citellus citellus) from Yugoslavia (Rodentia, Sciuridae). Italian Journal of Zoology, 57(4), 351-355.

\item[] Markov, G., Heltai, M., Nikolov, I., Penezić, A., Lanszki, J., \& Ćirović, D. (2018). Epigenetic variation and distinctness of golden jackal (Canis aureus) populations in its expanding Southeast European range. Comptes rendus de l'Academie bulgare des Sciences, 71(6), 787-793.


\item[] McGrath, J. W., Cheverud J. M., \&  Buikstra J. E. (1984). Genetic correlations between sides and heritability of asymmetry for nonmetric traits in rhesus macaques on Cayo Santiago. Am~J~Phys~Anthropol, 64(4) 401-411.


\item[] Meza-Pe\~naloza, A., \& Zertuche, F. (2019, April). Comparison by Non-Metric Traits of Xaltocan’s Shrine vs. Teotihuacan in Mexico. In C.~Morehart, \& C.~Frederick (Chairs), The Legacies of the Basin of Mexico: The Ecological Processes in the Evolution of a Civilization. Symposium conducted at the $ 84^{th} $~Annual Meeting of the Society for American Archaeology. Albuquerque N.~M.

\item[] Movsesian, A. A. (2013). Nonmetric Cranial Trait Variation and Population History of Medieval East Slavic Tribes. Am~J~Phys~Anthropol, 152, 495-505.

\item[] Movsesian, A. A., Bakholdina, V. Y., \& Pezhemsky, D. V. (2014). Biological Diversity and Population History of Middle Holocene Hunter-Gatherers from the Cis-Baikal Region of Siberia. Am~J~Phys~Anthropol, 155, 559-570.

\item[] Nikita, E. (2015). A Critical Review of the Mean Measure of Divergence and Mahalanobis Distances Using Artificial Data and New Approaches to the Estimation of Biodistances Employing Nonmetric Traits. Am~J~Phys~Anthropol, 157(2), 284-294.

\item[]  Ossenberg, N. S. (1976). Within and between race distances in population studies based on discrete traits of the human skull. Am~J~Phys~Anthropol, 45(3), 701-715.

\item[] Pink, C., Maier, C., Pilloud M., \& Hefner, J. (2016). Cranial nonmetric and morphoscopic data sets. In M. A. Pilloud, \& J. T. Hefner (Eds.), Biological Distance Analysis: Forensic and Bioarchaeological Perspectives (pp.~91-107). Elsevier, Academic Press. doi.org/10.1016/B978-0-12-801966-5.12001-3

\item[] Press, W.H., Teukolsky, S.A., Vetterling, W.T., \& Flannery, B.P. (1996). Numerical Recipes in C. (Ed.). Cambridge University press Cambridge, Cambridge.

\item[] Rothhammer, F., Quevedo, S., Cocilovo, J. A., \& Llop, E. (1984). Microevolution in prehistoric Andean populations: II. Chronologic nonmetrical cranial variation in northern Chile. Am~J~Phys~Anthropol, 65(2), 157-162.


\item[] Schillaci, M. A., Irish, J. D., \& Carolan C.E. Wood, C. C. E. (2009). Further Analysis of the Population History of Ancient Egyptians. \break Am~J~Phys~Anthropol, 139, 235-243.

\item[] Shapiro, S. S., \& Wilk, M. B. (1965). An analysis of variance test for normality (complete samples). Biometrika 52(3-4), 591-611.
    
\item[] Sikorski, M. D. (1982). Non-metrical divergence of isolated populations of Apodemus agrarius in urban areas. Acta theriol, 27(13), 169-180.

\item[] Sj{\o}vold, T. (1977). Non-metrical divergence between skeletal populations: the theoretical foundation and biological importance of CAB Smith's mean measure of divergence. Ossa, 4(1), 1-133.

\item[] de Souza, P., \& Houghton, P. (1977). The mean measure of divergence and the use of non-metric data in the estimation of biological distances. J~Archaeol~Sci, 4(2), 163-169.

\item[] Suryakan, \& Mahara T. (2016). A New Similarity Measure Based on Mean Measure of Divergence for Collaborative Filtering in Sparse Environment. Procedia Computer Science 89, 450-456.

\item[] Sutter, R. C., \& Mertz, L. (2004). Nonmetric cranial trait variation and prehistoric biocultural change in the Azapa Valley, Chile. American Journal of Physical Anthropology: The Official Publication of the American Association of Physical Anthropologists, 123(2), 130-145.

\item[]Trimble, M., \& Praderi, R. (2008). Assessment of nonmetric skull characters of the Franciscana (Pontoporia blainvillei) in determining population differences. Aquatic Mammals, 34(3), 338.

\item[] Verna, E., Piercecchi-Marti, M.D., Chaumoitre, K., Bartoli, C., Leonetti, G., \& Adalian, P. (2013). Discrete traits of the sternum and ribs: a useful contribution to identification in forensic anthropology and medicine. J~Forensic~Sci, 58(3), 571-577.


\item[] Villmoare, B. (2005). Metric and non-metric randomization methods, geographic variation, and the single-species hypothesis for Asian and African {\it Homo erectus}. J~Hum~Evol, 49, 680-701.

\item[] Voisin, J. L. (2012). Les caract\`{e}res discrets des membres sup\'{e}rieurs: un essai de synth\`{e}se des donn\'{e}es. Bulletins et m\'{e}moires de la Soci\'{e}t\'{e} d'anthropologie de Paris, 24(3-4), 107-130.

\item[] Wackerly, D. D., Mendenhall, M., \& Scheaffer, R. L. (2008). Mathematical Statistics with Applications. Thomson Higher Education. Pages 271 \& 274.

\item[] Williams, F. L. E., \& Cofran, Z. (2016). Postnatal craniofacial ontogeny in neandertals and modern humans. American journal of physical anthropology, 159(3), 394-409.

\end{itemize}

\newpage

\centerline{{\bf Figures}}

\begin{center}

\includegraphics[scale= 0.9]{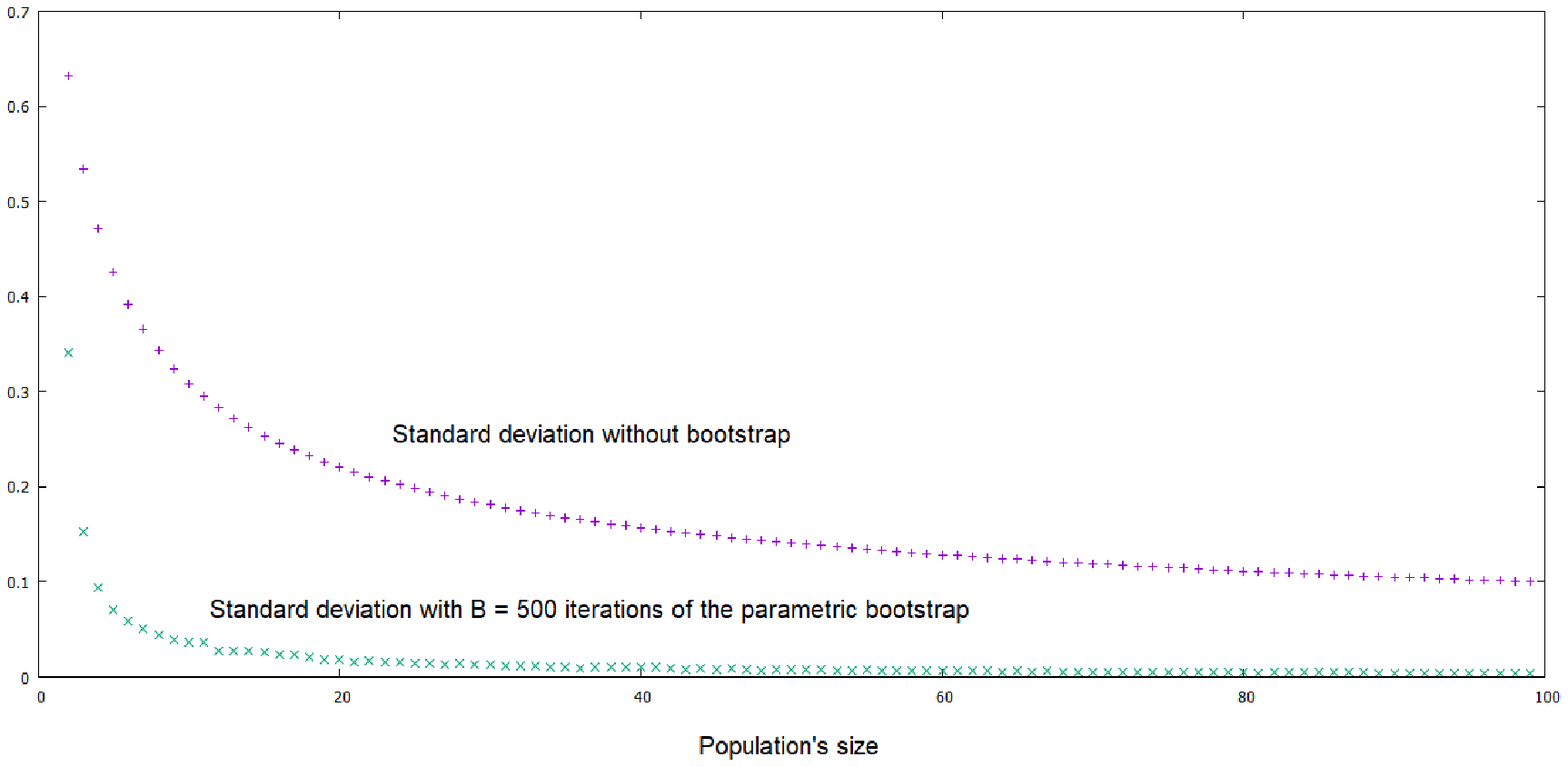}

\end{center}

Figure 1. Comparative graph of the standard deviations $ \sigma $ and the parametric bootstrapped $ \mathfrak{S} $ {\it vs.} the population's size.

\newpage

\begin{center}

\includegraphics[scale= 0.9]{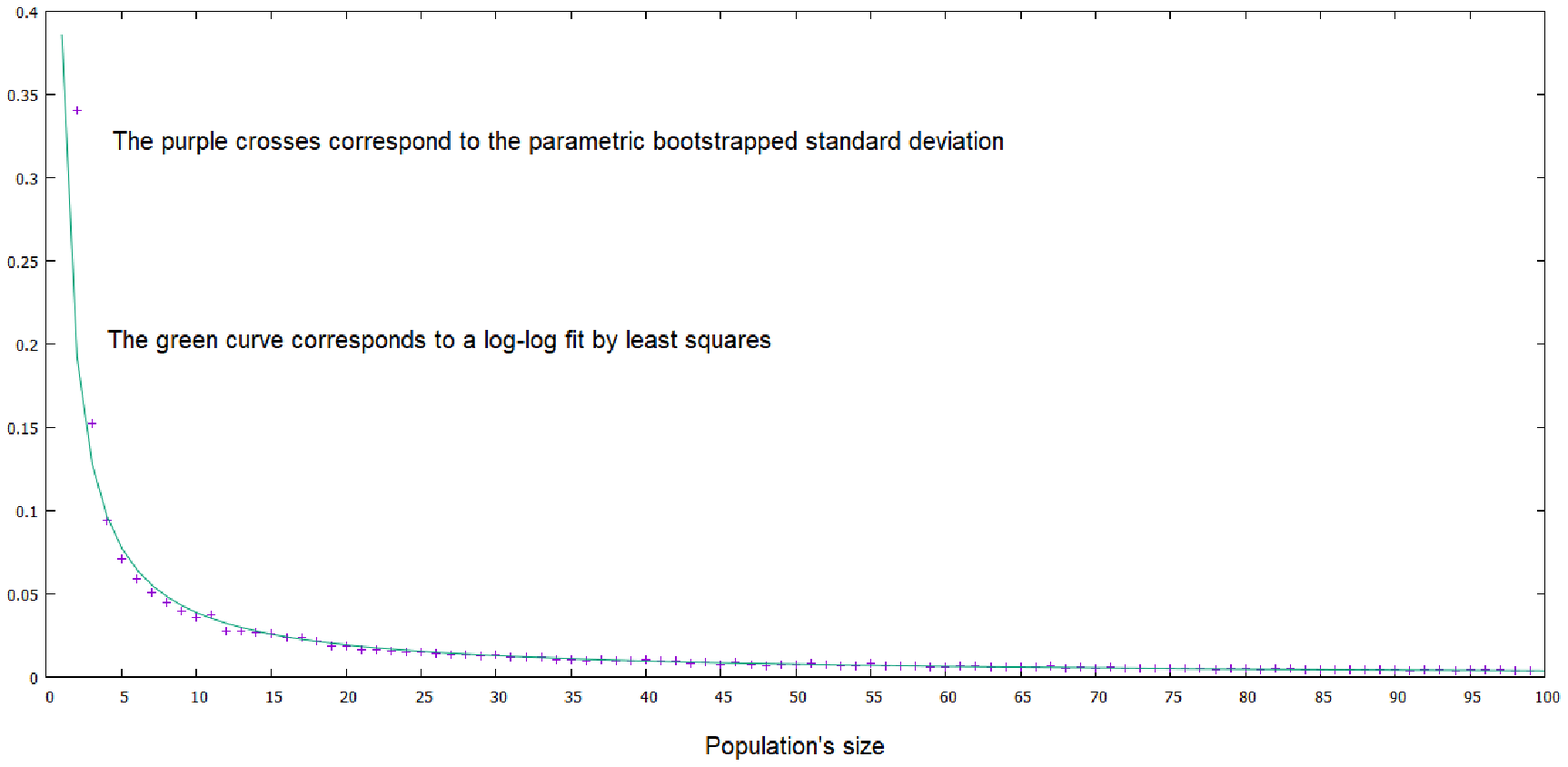}

\end{center}

Figure 2. The parametric bootstrapped $ \mathfrak{S} $ (in purple crosses) {\it vs.} the the population's size in cross dots. A log-log fit by least squares is shown in green.

\newpage

\begin{center}

\includegraphics{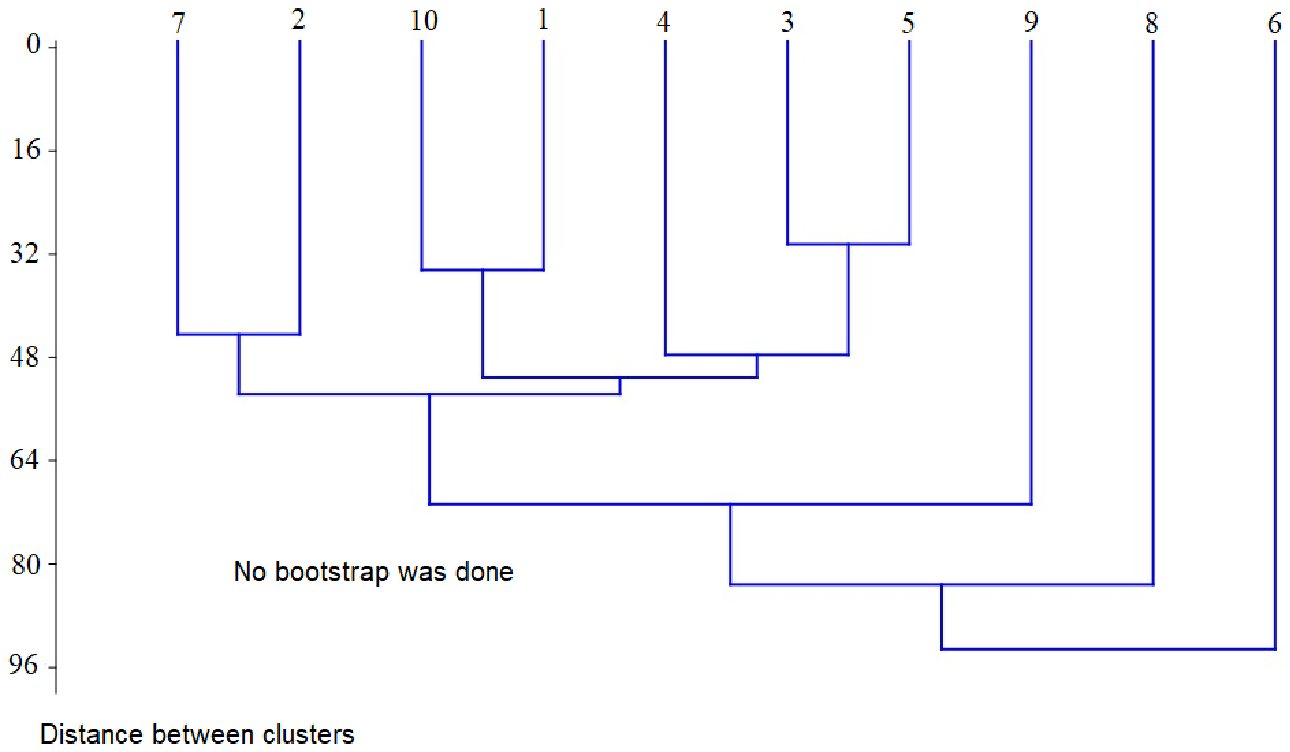}

\end{center}

Figure 3. The UPGMA dendrogram for the ten artificial populations of the distances matrix S2. No bootstrap was done.

\newpage

\begin{center}

\includegraphics{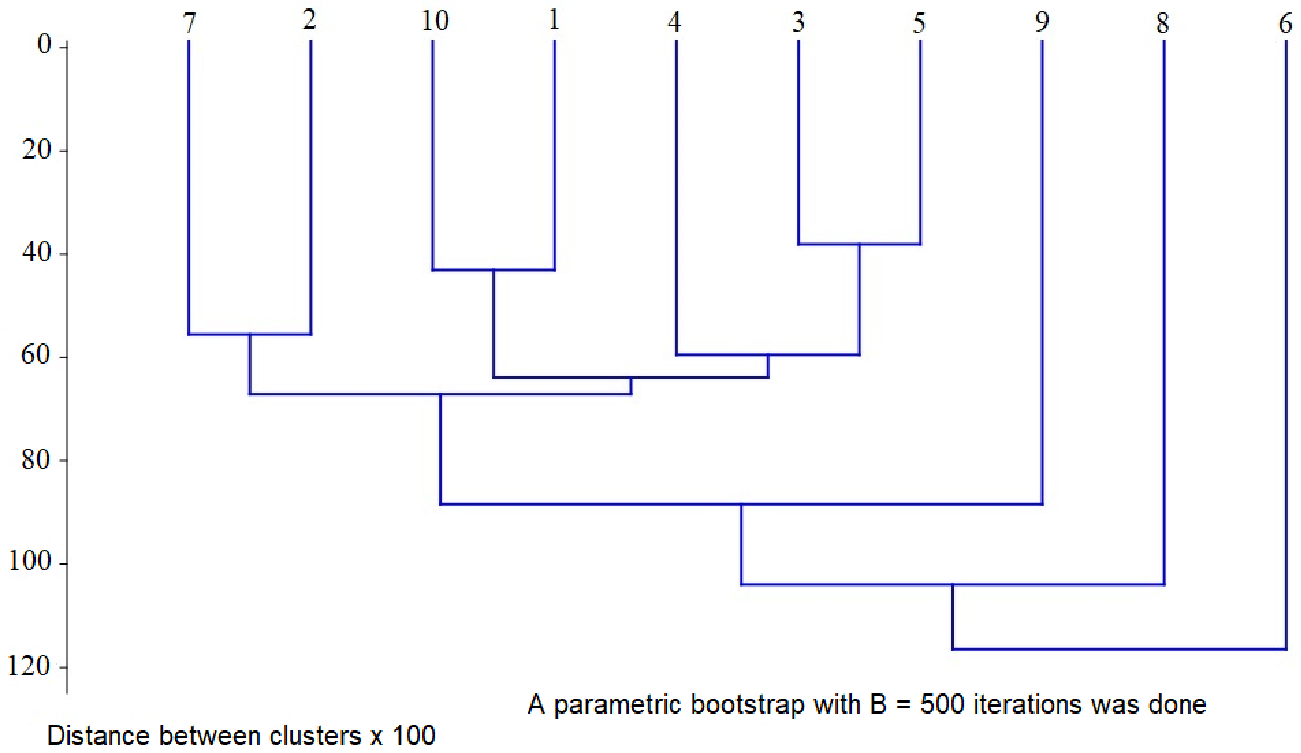}

\end{center}

Figure 4. The UPGMA dendrogram for the ten artificial populations of the distances matrix S3. A parametric bootstrap with $ B = 500 $ iterations was done.

\newpage

\begin{center}

\includegraphics{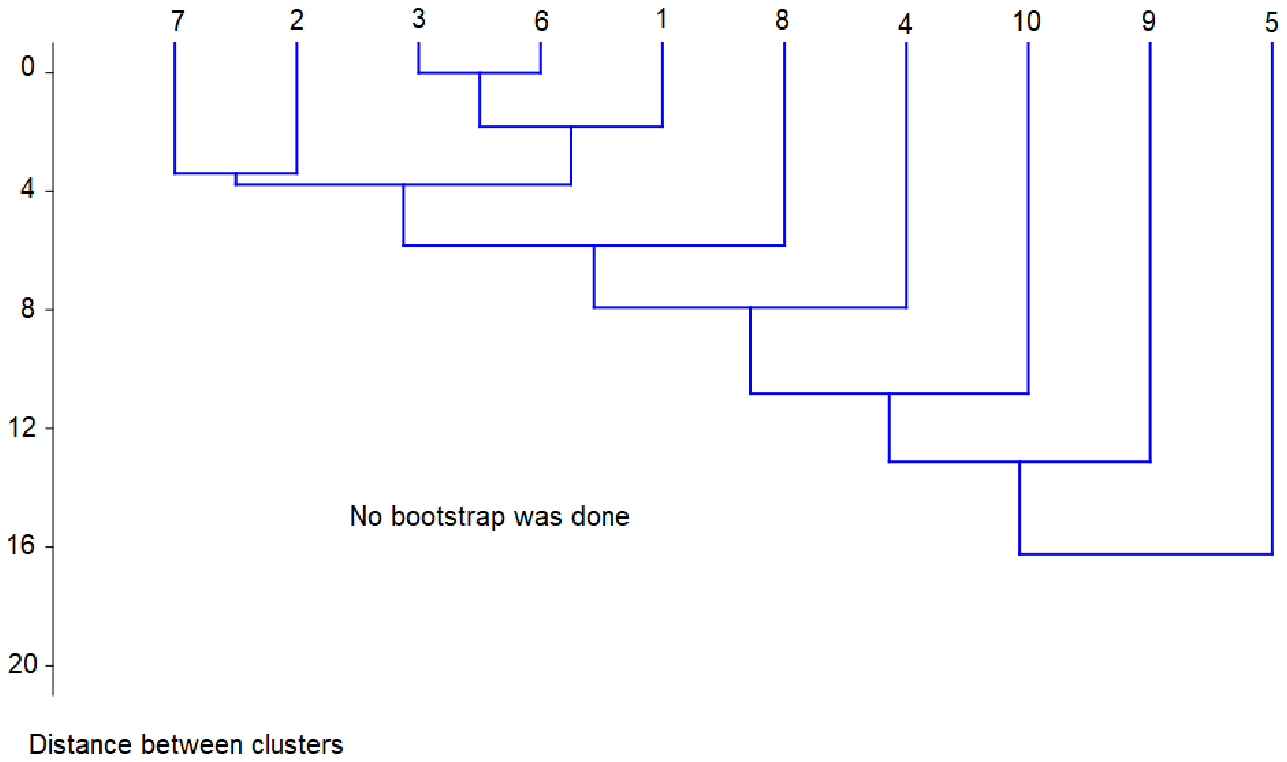}

\end{center}

Figure 5. The UPGMA dendrogram for the ten artificial populations of the distances matrix S5. No bootstrap was done.

\newpage

\begin{center}

\includegraphics{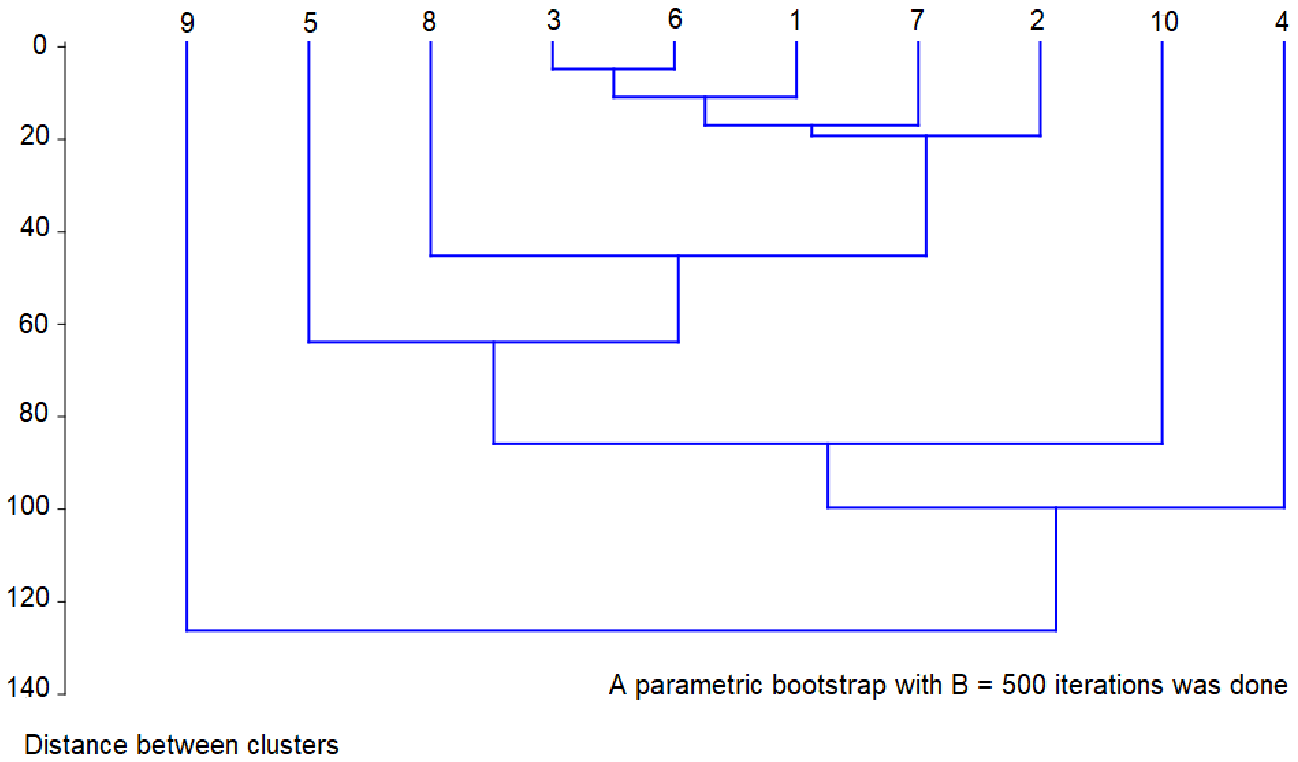}

\end{center}

Figure 6. The UPGMA dendrogram for the ten artificial populations of the distances matrix S6. A parametric bootstrap with $ B = 500 $ iterations was done.

\newpage

\begin{center}

\includegraphics{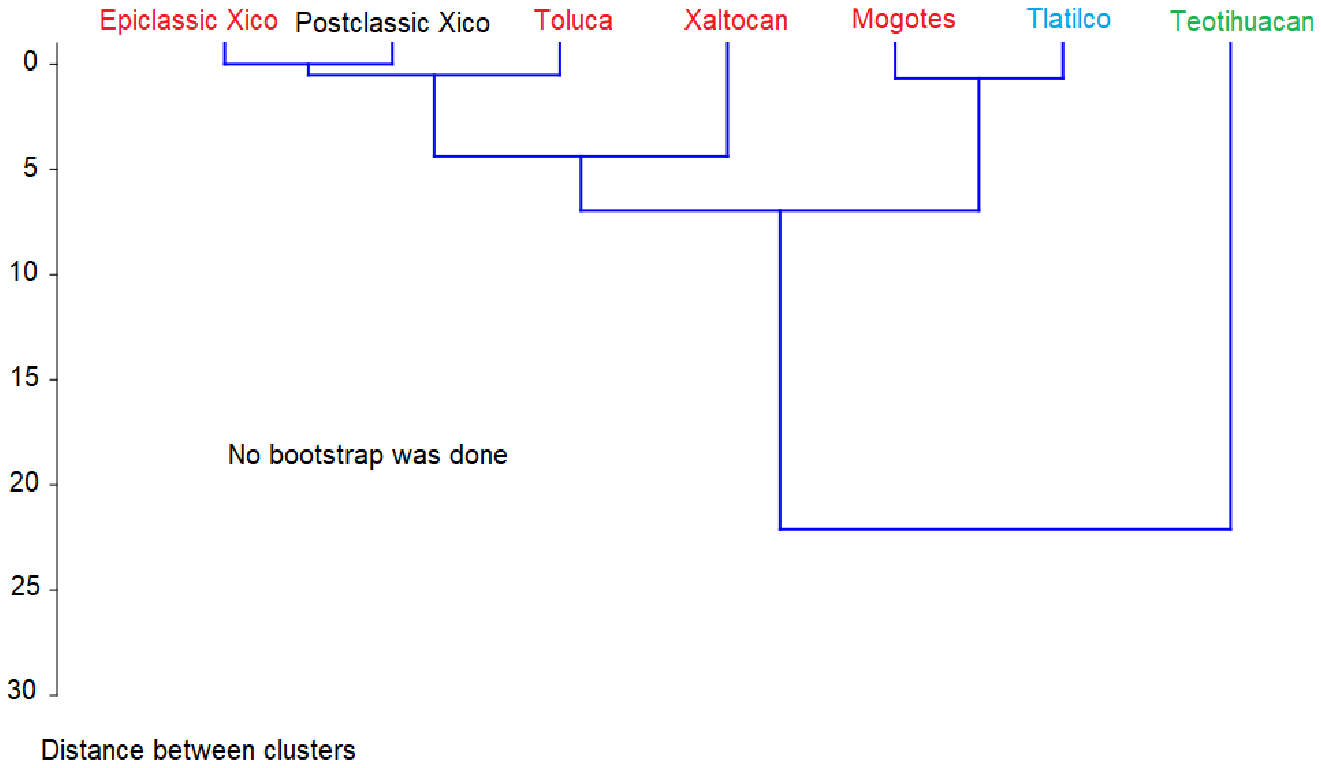}

\end{center}

Figure 7. The UPGMA dendrogram for the seven real populations of the distances matrix S8. No bootstrap was done. Colors indicate periods: Blue corresponds to {\it Preclassic}, green to {\it Classic}, red to {\it Epiclassic}, and black to {\it Postclassic}.

\newpage

\begin{center}

\includegraphics{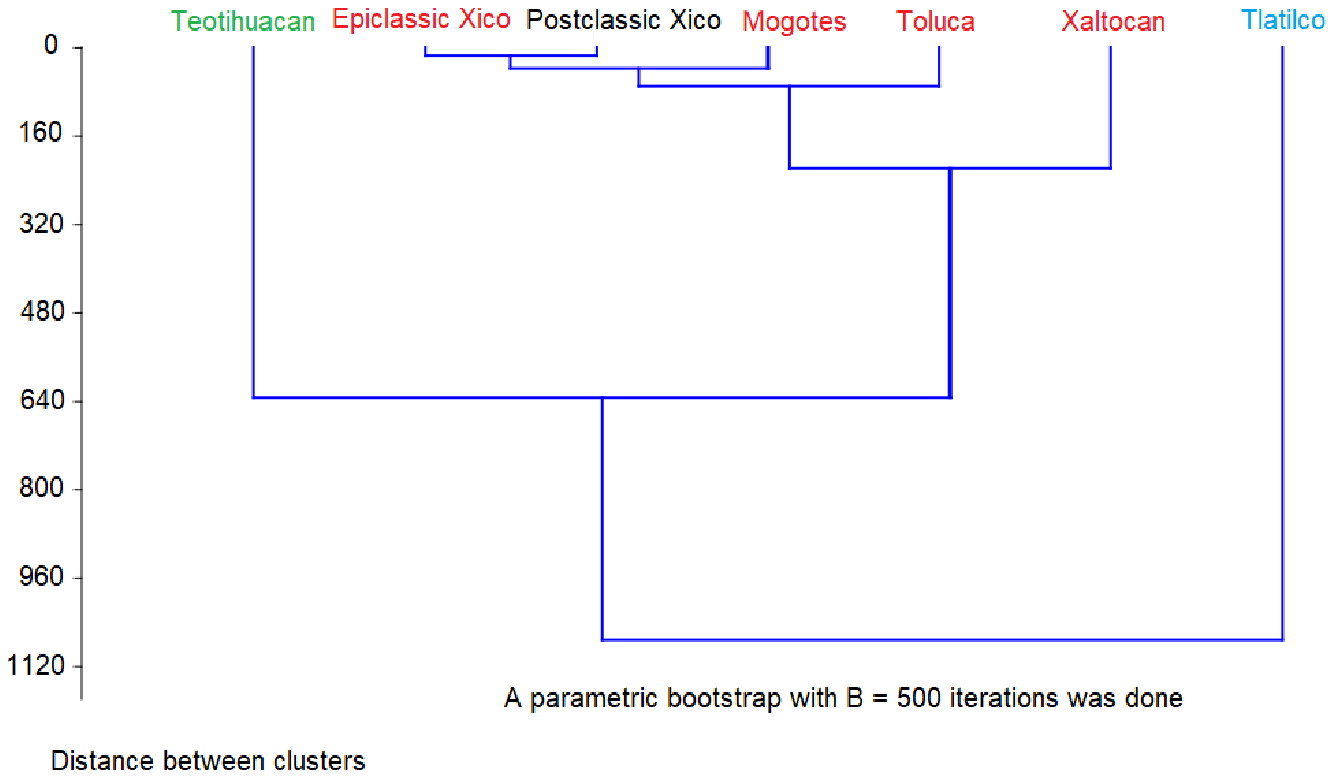}

\end{center}

Figure 8. The UPGMA dendrogram for the seven real populations of the distances matrix S9. A parametric bootstrap with $ B = 500 $ iterations was done. Colors indicate periods: Blue corresponds to {\it Preclassic}, green to {\it Classic}, red to {\it Epiclassic}, and black to {\it Postclassic}.

\newpage

\centerline{{\bf Supplementary Material}}

\

\footnotesize
\begin{center}
\begin{tabular}{|c|c|c|c|c|c|c|c|c|c|c|c|c|c|}
\hline $ \textrm{Population} \ \mu \downarrow \ \backslash \ i-\textrm{trait} \rightarrow $ & 1 & 2 & 3 & 4 & 5 & 6 & 7 & 8 & 9 & 10 & 11 & 12 & 13 \\
\hline $ N_1 = 93 $	&	&	&	&	&	&	&	&	&	&	&	&	&	\\
\hline $ N_{1i} \longrightarrow $	& 82	& 89	& 54	& 63	& 79	& 58	& 92	& 49	& 74	& 75	& 86	& 78	& 65	\\
\hline $ K_{1i} \longrightarrow $	& 58	& 46	& 10	& 54	& 9	& 55	& 77	& 8	& 40	& 26	& 8	& 33	& 64	\\
\hline $ N_2 = 76 $	&	&	&	&	&	&	&	&	&	&	&	&	&	\\
\hline $ N_{2i} \longrightarrow $	& 74	& 46	& 56	& 61	& 53	& 54	& 74	& 64	& 75	& 70	& 71	& 56	& 52	\\
\hline $ K_{2i} \longrightarrow $	& 52	& 44	& 47	& 36	& 8	& 16	& 68	& 44	& 24	& 46	& 32	& 22	& 21	\\
\hline $ N_3 = 68 $	&	&	&	&	&	&	&	&	&	&	&	&	&	\\
\hline $ N_{3i} \longrightarrow $	& 59	& 48	& 50	& 56	& 55	& 68	& 58	& 66	& 55	& 64	& 50	& 45	& 58	\\
\hline $ K_{3i} \longrightarrow $	& 6	& 37	& 6	& 13	& 13	& 5	& 55	& 4	& 43	& 32	& 8	& 13	& 43	\\
\hline $ N_4 = 58 $	&	&	&	&	&	&	&	&	&	&	&	&	&	\\
\hline $ N_{4i} \longrightarrow $	& 55	& 46	& 49	& 54	& 55	& 55	& 55	& 53	& 57	& 54	& 49	& 56	& 45	\\
\hline $ K_{4i} \longrightarrow $	& 12	& 10	& 31	& 36	& 21	& 41	& 7	& 9	& 22	& 36	& 16	& 48	& 26	\\
\hline $ N_5 = 64 $	&	&	&	&	&	&	&	&	&	&	&	&	&	\\
\hline $ N_{5i} \longrightarrow $	& 59	& 61	& 58	& 64	& 56	& 62	& 61	& 53	& 52	& 52	& 45	& 64	& 45	\\
\hline $ K_{5i} \longrightarrow $	& 36	& 24	& 19	& 42	& 26	& 12	& 35	& 4	& 43	& 19	& 33	& 31	& 41	\\
\hline $ N_6 = 64 $	&	&	&	&	&	&	&	&	&	&	&	&	&	\\
\hline $ N_{6i} \longrightarrow $	& 46	& 53	& 62	& 62	& 46	& 64	& 58	& 57	& 53	& 46	& 46	& 61	& 58	\\
\hline $ K_{6i} \longrightarrow $	& 5	& 35	& 6	& 3	& 32	& 7	& 1	& 49	& 9	& 44	& 21	& 11	& 39	\\
\hline $ N_7 = 93 $	&	&	&	&	&	&	&	&	&	&	&	&	&	\\
\hline $ N_{7i} \longrightarrow $	& 61	& 79	& 51	& 55	& 65	& 89	& 82	& 64	& 57	& 83	& 65	& 79	& 66	\\
\hline $ K_{7i} \longrightarrow $	& 34	& 74	& 17	& 31	& 41	& 82	& 68	& 22	& 43	& 75	& 35	& 11	& 59	\\
\hline $ N_8 = 62 $	&	&	&	&	&	&	&	&	&	&	&	&	&	\\
\hline $ N_{8i} \longrightarrow $	& 46	& 46	& 58	& 59	& 59	& 52	& 52	& 58	& 47	& 56	& 57	& 46	& 58	\\
\hline $ K_{8i} \longrightarrow $	& 30	& 28	& 56	& 58	& 45	& 52	& 41	& 49	& 37	& 13	& 6	& 46	& 10	\\
\hline $ N_9 = 72 $	&	&	&	&	&	&	&	&	&	&	&	&	&	\\
\hline $ N_{9i} \longrightarrow $	& 57	& 57	& 66	& 64	& 59	& 72	& 67	& 70	& 48	& 52	& 68	& 72	& 62	\\
\hline $ K_{9i} \longrightarrow $	& 32	& 22	& 18	& 3	& 5	& 58	& 16	& 44	& 48	& 42	& 12	& 43	& 4	\\
\hline $ N_{10} = 85 $	&	&	&	&	&	&	&	&	&	&	&	&	&	\\
\hline $ N_{10i} \longrightarrow $	& 65	& 47	& 58	& 51	& 51	& 57	& 48	& 75	& 46	& 78	& 50	& 62	& 77	\\
\hline $ K_{10i} \longrightarrow $	& 59	& 11	& 2	& 38	& 25	& 46	& 42	& 40	& 33	& 20	& 5	& 35	& 19	\\
\hline

\end{tabular}
\end{center}
\noindent
\textbf{Table~S1.} The ten populations of the first artificial data generated randomly with their sizes and frequencies for each trait.

\normalsize
\baselineskip = 24.8pt

\newpage

\footnotesize
\begin{center}
\begin{tabular}{|c|c|c|c|c|c|c|c|c|c|c|}
\hline $  \mu \downarrow \ \backslash \ \nu \rightarrow $ & {\bf 1} & {\bf 2} & {\bf 3} & {\bf 4} & {\bf 5} & {\bf 6} & {\bf 7} & {\bf 8} & {\bf 9} & {\bf 10}  \\
\hline {\bf 1}	& 0.00& 64.36& 56.45& 48.48& 40.21& 127.80& 40.60& 84.19& 92.83& 34.48\\
\hline {\bf 2}	& 64.36& 0.00& 51.41& 57.71& 50.07& 80.15& 44.44& 67.69& 75.31& 59.56\\
\hline {\bf 3}	& 56.45& 51.41& 0.00& 61.84& 30.44& 70.97& 51.94& 128.26& 73.61& 62.91\\
\hline {\bf 4}	& 48.48& 57.71& 61.84& 0.00& 33.31& 61.65& 59.41& 57.70& 50.16& 52.11\\
\hline {\bf 5}	& 40.21& 50.07& 30.44& 33.31& 0.00& 74.32& 41.75& 93.37& 73.68& 46.43\\
\hline {\bf 6}	& 127.80& 80.15& 70.97& 61.65& 74.32& 0.00& 79.21& 159.14& 77.84& 107.45\\
\hline {\bf 7}	& 40.60& 44.44& 51.94& 59.41& 41.75& 79.21& 0.00& 91.57& 78.25& 59.65\\
\hline {\bf 8}	& 84.19& 67.69& 128.26& 57.70& 93.37& 159.14& 91.57& 0.00& 90.66& 51.72\\
\hline {\bf 9}	& 92.83& 75.31& 73.61& 50.16& 73.68& 77.84& 78.25& 90.66& 0.00& 51.16\\
\hline {\bf 10}	& 34.48& 59.56& 62.91& 52.11& 46.43& 107.45& 59.65& 51.72& 51.16& 0.00\\
\hline

\end{tabular}
\end{center}
\noindent
\textbf{Table~S2.} The distances matrix for the ten populations in Table~S1. No parametric bootstrap was done.

\normalsize

\baselineskip = 24.8pt

\footnotesize
\begin{center}
\begin{tabular}{|c|c|c|c|c|c|c|c|}
\hline $  \mu \downarrow \ \backslash \ \nu \rightarrow $ & {\bf 1} & {\bf 2} & {\bf 3} & {\bf 4} & {\bf 5} & {\bf 6} & {\bf 7} \\
\hline {\bf 1}	& 0.00& 7981.11& 6932.59& 5643.70& 4880.85& 14770.31& 5448.63\\
\hline {\bf 2}	& 7981.11& 0.00& 5975.68& 6359.35& 5742.36& 8928.15& 5650.30\\
\hline {\bf 3}	& 6932.59& 5975.68& 0.00& 6728.24& 3549.22& 7819.99& 6377.81\\
\hline {\bf 4}	& 5643.70& 6359.35& 6728.24& 0.00& 3661.09& 6517.30& 6866.39\\
\hline {\bf 5}	& 4880.85& 5742.36& 3549.22& 3661.09& 0.00& 7941.31& 5062.57\\
\hline {\bf 6}	& 14770.31& 8928.15& 7819.99& 6517.30& 7941.31& 0.00& 9187.00\\
\hline {\bf 7}	& 5448.63& 5650.30& 6377.81& 6866.39& 5062.57& 9187.00& 0.00\\
\hline {\bf 8}	& 9749.12& 7507.95& 13733.05& 6098.67& 9937.99& 16526.58& 10564.24\\
\hline {\bf 9}	& 12004.17& 9203.88& 8818.96& 5783.09& 8583.90& 8874.16& 10066.34\\
\hline {\bf 10}	& 4292.37& 6789.54& 7052.97& 5645.17& 5182.53& 11457.20& 7129.90\\
\hline
\end{tabular}
\end{center}

\footnotesize
\begin{center}
\begin{tabular}{|c|c|c|c|}
\hline $  \mu \downarrow \ \backslash \ \nu \rightarrow $ & {\bf 8} & {\bf 9} & {\bf 10} \\
\hline {\bf 1}	& 9749.12& 12004.17& 4292.37\\
\hline {\bf 2}	& 7507.95& 9203.88& 6789.54\\
\hline {\bf 3}	& 13733.05& 8818.96& 7052.97\\
\hline {\bf 4}	& 6098.67& 5783.09& 5645.17\\
\hline {\bf 5}	& 9937.99& 8583.90& 5182.53\\
\hline {\bf 6}	& 16526.58& 8874.16& 11457.20\\
\hline {\bf 7}	& 10564.24& 10066.34& 7129.90\\
\hline {\bf 8}	& 0.00& 10263.92& 5624.98\\
\hline {\bf 9}	& 10263.92& 0.00& 6076.19\\
\hline {\bf 10}	& 5624.98& 6076.19& 0.00\\
\hline
\end{tabular}
\end{center}
\noindent
\textbf{Table~S3.} The distances matrix for the ten populations in Table~S1 after a parametric bootstrap with $ B = 500 $ iterations.

\normalsize

\baselineskip = 24.8pt

\footnotesize
\begin{center}
\begin{tabular}{|c|c|c|c|c|c|c|c|c|c|c|c|c|c|}
\hline $ \textrm{Population} \ \mu \downarrow \ \backslash \ i-\textrm{trait} \rightarrow $ & 1 & 2 & 3 & 4 & 5 & 6 & 7 & 8 & 9 & 10 & 11 & 12 & 13 \\
\hline $ N_{1} = 46 $	&	&	&	&	&	&	&	&	&	&	&	&	&	\\
\hline $ N_{1i} \longrightarrow $	& 4	& 2	& 7	& 37	& 2	& 9	& 10	& 15	& 38	& 35	& 4	& 28	& 16	\\
\hline $ K_{1i} \longrightarrow $	& 3	& 0	& 3	& 16	& 2	& 4	& 2	& 2	& 4	& 8	& 4	& 7	& 1	\\
\hline $ N_{2} = 76 $	&	&	&	&	&	&	&	&	&	&	&	&	&	\\
\hline $ N_{2i} \longrightarrow $	& 75	& 32	& 6	& 6	& 44	& 71	& 45	& 68	& 75	& 72	& 21	& 33	& 62	\\
\hline $ K_{2i} \longrightarrow $	& 36	& 14	& 3	& 1	& 42	& 38	& 15	& 33	& 29	& 16	& 13	& 21	& 59	\\
\hline $ N_{3} = 4 $	&	&	&	&	&	&	&	&	&	&	&	&	&	\\
\hline $ N_{3i} \longrightarrow $	& 4	& 3	& 2	& 2	& 2	& 3	& 2	& 4	& 2	& 2	& 2	& 4	& 4	\\
\hline $ K_{3i} \longrightarrow $	& 4	& 0	& 1	& 1	& 0	& 2	& 2	& 1	& 1	& 0	& 1	& 4	& 2	\\
\hline $ N_{4} = 59 $	&	&	&	&	&	&	&	&	&	&	&	&	&	\\
\hline $ N_{4i} \longrightarrow $	& 43	& 6	& 26	& 31	& 27	& 14	& 40	& 48	& 29	& 20	& 27	& 4	& 56	\\
\hline $ K_{4i} \longrightarrow $	& 38	& 3	& 18	& 8	& 0	& 13	& 9	& 7	& 25	& 11	& 26	& 3	& 45	\\
\hline $ N_{5} = 99 $	&	&	&	&	&	&	&	&	&	&	&	&	&	\\
\hline $ N_{5i} \longrightarrow $	& 3	& 54	& 43	& 69	& 78	& 17	& 72	& 28	& 37	& 63	& 83	& 75	& 40	\\
\hline $ K_{5i} \longrightarrow $	& 0	& 16	& 11	& 22	& 17	& 0	& 65	& 26	& 0	& 14	& 37	& 39	& 38	\\
\hline $ N_{6} = 5 $	&	&	&	&	&	&	&	&	&	&	&	&	&	\\
\hline $ N_{6i} \longrightarrow $	& 5	& 5	& 3	& 2	& 2	& 4	& 3	& 4	& 4	& 3	& 3	& 4	& 4	\\
\hline $ K_{6i} \longrightarrow $	& 1	& 0	& 3	& 1	& 0	& 4	& 2	& 3	& 1	& 0	& 3	& 3	& 0	\\
\hline $ N_{7} = 66 $	&	&	&	&	&	&	&	&	&	&	&	&	&	\\
\hline $ N_{7i} \longrightarrow $	& 43	& 8	& 2	& 6	& 37	& 27	& 28	& 6	& 63	& 3	& 51	& 37	& 22	\\
\hline $ K_{7i} \longrightarrow $	& 43	& 2	& 1	& 3	& 21	& 20	& 1	& 4	& 11	& 0	& 24	& 9	& 21	\\
\hline $ N_{8} = 36 $	&	&	&	&	&	&	&	&	&	&	&	&	&	\\
\hline $ N_{8i} \longrightarrow $	& 20	& 11	& 20	& 6	& 16	& 24	& 6	& 35	& 25	& 26	& 31	& 8	& 23	\\
\hline $ K_{8i} \longrightarrow $	& 12	& 1	& 4	& 4	& 7	& 19	& 3	& 23	& 25	& 3	& 14	& 4	& 23	\\
\hline $ N_{9} = 88 $	&	&	&	&	&	&	&	&	&	&	&	&	&	\\
\hline $ N_{9i} \longrightarrow $	& 65	& 74	& 5	& 70	& 84	& 62	& 20	& 8	& 31	& 86	& 41	& 38	& 75	\\
\hline $ K_{9i} \longrightarrow $	& 40	& 70	& 0	& 28	& 27	& 4	& 5	& 2	& 21	& 69	& 29	& 28	& 32	\\
\hline $ N_{10} = 38 $	&	&	&	&	&	&	&	&	&	&	&	&	&	\\
\hline $ N_{10i} \longrightarrow $	& 32	& 10	& 33	& 16	& 35	& 29	& 20	& 22	& 26	& 4	& 18	& 18	& 38	\\
\hline $ K_{10i} \longrightarrow $	& 4	& 0	& 7	& 1	& 20	& 26	& 15	& 11	& 21	& 4	& 2	& 7	& 23	\\
\hline

\end{tabular}
\end{center}
\noindent
\textbf{Table~S4.} The ten populations of the second artificial data generated randomly with their sizes and frequencies for each trait.

\normalsize

\baselineskip = 24.8pt

\newpage

\footnotesize
\begin{center}
\begin{tabular}{|c|c|c|c|c|c|c|c|c|c|c|}
\hline $  \mu \downarrow \ \backslash \ \nu \rightarrow $ & {\bf 1} & {\bf 2} & {\bf 3} & {\bf 4} & {\bf 5} & {\bf 6} & {\bf 7} & {\bf 8} & {\bf 9} & {\bf 10}  \\
\hline {\bf 1}	& 0.00& 4.68& 1.71& 8.89& 11.99& 1.93& 3.87& 9.92& 8.41& 10.71\\
\hline {\bf 2}	& 4.68& 0.00& 2.26& 15.17& 15.22& 5.28& 3.39& 7.57& 14.73& 14.16\\
\hline {\bf 3}	& 1.71& 2.26& 0.00& 1.22& 4.18& 0.00& 1.71& 1.11& 3.79& 3.54\\
\hline {\bf 4}	& 8.89& 15.17& 1.22& 0.00& 31.05& 4.26& 8.12& 9.85& 14.72& 18.02\\
\hline {\bf 5}	& 11.99& 15.22& 4.18& 31.05& 0.00& 7.04& 13.49& 18.94& 22.88& 21.39\\
\hline {\bf 6}	& 1.93& 5.28& 0.00& 4.26& 7.04& 0.00& 4.78& 5.25& 9.38& 6.05\\
\hline {\bf 7}	& 3.87& 3.39& 1.71& 8.12& 13.49& 4.78& 0.00& 5.30& 12.11& 13.13\\
\hline {\bf 8}	& 9.92& 7.57& 1.11& 9.85& 18.94& 5.25& 5.30& 0.00& 19.36& 10.09\\
\hline {\bf 9}	& 8.41& 14.73& 3.79& 14.72& 22.88& 9.38& 12.11& 19.36& 0.00& 22.52\\
\hline {\bf 10}	& 10.71& 14.16& 3.54& 18.02& 21.39& 6.05& 13.13& 10.09& 22.52& 0.00\\
\hline
\end{tabular}
\end{center}
\noindent
\textbf{Table~S5.} The distances matrix for the ten populations in Table~S4. No bootstrap was done.

\normalsize
\baselineskip = 24.8pt

\

\footnotesize
\begin{center}
\begin{tabular}{|c|c|c|c|c|c|c|c|c|c|c|}
\hline $  \mu \downarrow \ \backslash \ \nu \rightarrow $ & {\bf 1} & {\bf 2} & {\bf 3} & {\bf 4} & {\bf 5} & {\bf 6} & {\bf 7} & {\bf 8} & {\bf 9} & {\bf 10}  \\
\hline {\bf 1}	& 0.00& 20.17& 9.74& 35.42& 43.30& 12.06& 17.94& 40.31& 31.74& 41.46\\
\hline {\bf 2}	& 20.17& 0.00& 11.68& 174.08& 98.28& 26.41& 18.60& 119.79& 169.60& 169.74\\
\hline {\bf 3}	& 9.74& 11.68& 0.00& 9.05& 17.14& 4.78& 9.76& 8.90& 16.09& 16.02\\
\hline {\bf 4}	& 35.42& 174.08& 9.05& 0.00& 197.95& 23.84& 38.51& 125.62& 161.67& 192.50\\
\hline {\bf 5}	& 43.30& 98.28& 17.14& 197.95& 0.00& 32.48& 53.39& 138.46& 142.91& 140.73\\
\hline {\bf 6}	& 12.06& 26.41& 4.78& 23.84& 32.48& 0.00& 23.20& 28.19& 41.54& 30.76\\
\hline {\bf 7}	& 17.94& 18.60& 9.76& 38.51& 53.39& 23.20& 0.00& 28.78& 50.66& 58.71\\
\hline {\bf 8}	& 40.31& 119.79& 8.90& 125.62& 138.46& 28.19& 28.78& 0.00& 276.02& 143.70\\
\hline {\bf 9}	& 31.74& 169.60& 16.09& 161.67& 142.91& 41.54& 50.66& 276.02& 0.00& 245.79\\
\hline {\bf 10}	& 41.46& 169.74& 16.02& 192.50& 140.73& 30.76& 58.71& 143.70& 245.79& 0.00\\
\hline
\end{tabular}
\end{center}
\noindent
\textbf{Table~S6.} The distances matrix for the ten populations in Table~S4 after a parametric bootstrap with $ B = 500 $ iterations.

\normalsize

\baselineskip = 24.8pt

\footnotesize
\begin{center}
\begin{tabular}{|c|c|c|c|c|c|c|c|c|c|c|c|c|c|}
\hline $ \textrm{Population} \ \mu \downarrow \ \backslash \ i-\textrm{trait} \rightarrow $ & 1 & 2 & 3 & 4 & 5 & 6 & 7 & 8 & 9 & 10 & 11 & 12 & 13 \\
\hline \textrm{Tlatilco} $ N_1 = 78 $ &  &  &  &  &  &  & & & & & & &  \\
\hline $ N_{1i} \longrightarrow $  &77 &78 & 77 & 74 & 63 & 78 & 75 & 78 & 56 & 53 & 78 & 78 & 78  \\
\hline $ K_{1i} \longrightarrow $ & 3	& 52 &	14 &	10 &	14 &	66 &	34 &	11 &	3 &	5 &	9 &	16 &	43 \\
\hline \textrm{Teotihuacan} $ N_2 = 66 $  &  &  &  &  &  &  & & & & & & & \\
\hline $ N_{2i} \longrightarrow $ & 33 &	32 &	23 &	15 &	10 &	28 &	21 &	16 &	15 &	13 &	18 &	18 &	21	 \\
\hline $ K_{2i} \longrightarrow $ & 0 &	16 &	5 &	15 &	0 &	18 &	1 &	1 &	9 &	9 &	4 &	0 &	1 \\
\hline \textrm{Epiclassic Xico} $ N_3 = 5 $ &  &  &  &  &  &  & & & & & & &  \\
\hline $ N_{3i} \longrightarrow $ & 5 &	5&	5&	5&	5&	5&	4&	4&	3&	4&	5&	5&	5 \\
\hline $ K_{3i} \longrightarrow $ &0&	5&	4&	0&	2&	5&	3&	0&	2&	2&	4&	1&	1	 \\
\hline \textrm{Toluca} $ N_4 = 23 $&  &  &  &  &  &  & & & & & & &  \\
\hline $ N_{4i} \longrightarrow $  &23&	22&	21&	22&	21&	23&	20&	22&	21&	19&	23&	23&	23 \\
\hline $ K_{4i} \longrightarrow $  &0&	18&	9&	0&	8&	10&	15&	8&	10&	6&	9&	0&	6 \\
\hline \textrm{Xaltocan} $ N_5 = 118 $ &  &  &  &  &  &  & & & & & & &  \\
\hline $ N_{5i} \longrightarrow $ &93&	95&	82&	82&	80&	106&	94&	94&	95&	95&	116&	116&	116 \\
\hline $ K_{5i} \longrightarrow $ &0&	77&	41&	10&	49&	68&	65&	25&	8&	51&	32&	8&	80 \\
\hline \textrm{Mogotes}  $ N_6 = 15 $ &  &  &  &  &  &  & & & & & & &  \\
\hline $ N_{6i} \longrightarrow $ &15&	13&	8&	8&	11&	6&	8&	8&	12&	11&	13&	14&	11 \\
\hline $ K_{6i} \longrightarrow $ &0&	10&	3&	0&	5&	6&	5&	3&	1&	1&	3&	0&	4	 \\
\hline \textrm{Postclassic Xico} $ N_7 = 28 $ &  &  &  &  &  &  & & & & & & &  \\
\hline $ N_{7i} \longrightarrow $ &28&	28&	24&	24&	22&	27&	20&	19&	22&	22&	28&	28&	27	 \\
\hline $ K_{7i} \longrightarrow $ &0&	27&	9&	1&	9&	20&	14&	7&	7&	9&	10&	5&	9	\\
\hline

\end{tabular}
\end{center}
\noindent
\textbf{Table~S7.} Seven populations from the {\it Basin of Mexico} with their frequencies and their $13$ traits, corresponding to: $ 1 = $ Metopic suture, $ 2 = $ Supraorbital structures, $ 3 = $ Infraorbital suture, $ 4 = $ Multiple infraorbital foramina, $ 5 = $ Zygomatico facial foramina, $ 6 = $ Parietal Foramen, $ 7 = $ Condylar canal, $ 8 = $ Divided hypoglossal canal, $ 9 = $ Foramen Oval incomplete, $ 10 = $ Foramen Spinosum incomplete, $ 11 = $ Tympanic Dehiscence, $ 12 = $ Auditory exostosis, $ 13 = $ Mastoid Foramen.

\normalsize

\baselineskip = 24.8pt

\footnotesize
\begin{center}
\begin{tabular}{|c|c|c|c|c|c|c|c|}
\hline $  \mu \downarrow \ \backslash \ \nu \rightarrow $ & {\bf 1} & {\bf 2} & {\bf 3} & {\bf 4} & {\bf 5} & {\bf 6} & {\bf 7}  \\
\hline {\bf 1}	& 0.00& 28.84& 4.57& 13.55& 23.76& 0.68& 9.29\\
\hline {\bf 2}	& 28.84& 0.00& 8.14& 21.64& 36.46& 15.22& 22.41\\
\hline {\bf 3}	& 4.57& 8.14& 0.00& 0.75& 2.42& 1.07& 0.00\\
\hline {\bf 4}	& 13.55& 21.64& 0.75& 0.00& 6.75& 1.32& 0.29\\
\hline {\bf 5}	& 23.76& 36.46& 2.42& 6.75& 0.00& 1.84& 3.99\\
\hline {\bf 6}	& 0.68& 15.22& 1.07& 1.32& 1.84& 0.00& 0.24\\
\hline {\bf 7}	& 9.29& 22.41& 0.00& 0.29& 3.99& 0.24& 0.00\\
\hline

\end{tabular}
\end{center}
\noindent
\textbf{Table~S8.} The distances matrix for the seven populations in Table~S7. No bootstrap was done.

\normalsize

\baselineskip = 24.8pt

\footnotesize
\begin{center}
\begin{tabular}{|c|c|c|c|c|c|c|c|}
\hline $ \mu \downarrow \ \backslash \ \nu \rightarrow $ & {\bf 1} & {\bf 2} & {\bf 3}  & {\bf 4} & {\bf 5} & {\bf 6} & {\bf 7} \\
\hline {\bf 1}	& 0.00& 976.37& 45.08& 764.52& 3950.88& 50.74& 643.90\\
\hline {\bf 2}	& 976.37& 0.00& 78.30& 771.41& 1175.90& 326.84& 813.38\\
\hline {\bf 3}	& 45.08& 78.30& 0.00& 22.00& 30.25& 26.32& 14.23\\
\hline {\bf 4}	& 764.52& 771.41& 22.00& 0.00& 429.81& 69.73& 118.15\\
\hline {\bf 5}	& 3950.88& 1175.90& 30.25& 429.81& 0.00& 68.13& 346.69\\
\hline {\bf 6}	& 50.74& 326.84& 26.32& 69.73& 68.13& 0.00& 48.93\\
\hline {\bf 7}	& 643.90& 813.38& 14.23& 118.15& 346.69& 48.93& 0.00\\
\hline

\end{tabular}
\end{center}
\noindent
\textbf{Table~S9.} The distances matrix for the seven populations in Table~S7. A parametric bootstrap was applied with $ B = 500 $ iterations. All the distances are statistically representative at the $ \alpha \leq 10^{-17} $ level.

\normalsize

\baselineskip = 24.8pt

\end{document}